\documentclass{article}

\usepackage[english]{babel}
\usepackage[numbers]{natbib}

\usepackage[letterpaper,top=2cm,bottom=2cm,left=3cm,right=3cm,marginparwidth=1.75cm]{geometry}

\usepackage{amsmath}
\usepackage{booktabs}
\usepackage{graphicx}
\usepackage{tabularx}
\usepackage[colorlinks=true, allcolors=blue]{hyperref}

\newcolumntype{Y}{>{\raggedright\arraybackslash}X}

\title{AutoSR: Automatic Symbolic Regression by Searching Research States}
\author{%
Kejia Zhang\textsuperscript{1,*},
Youran Sun\textsuperscript{1,*},
Xinyu Ren\textsuperscript{2},
Chugang Yi\textsuperscript{1}, and
Haizhao Yang\textsuperscript{1,\textdagger}\\[0.5em]
\small \textsuperscript{1}Department of Mathematics, University of Maryland, College Park, MD, USA\\
\small \textsuperscript{2}The Chinese University of Hong Kong, Hong Kong SAR, China
}
\date{}

\begin{document}
\maketitle
\begingroup
\renewcommand{\thefootnote}{\fnsymbol{footnote}}
\footnotetext[1]{Kejia Zhang and Youran Sun contributed equally to this work.}
\footnotetext[2]{Corresponding author. Email: \href{mailto:hzyang@umd.edu}{hzyang@umd.edu}.}
\endgroup

\begin{abstract}
We introduce Automatic Symbolic Regression (AutoSR), a fully automated system that instantiates Research-Space Symbolic Regression by searching persistent scientific investigations rather than isolated equations.
Finite, noisy data often yield numerically competitive expressions that imply very different behavior outside the observed regime, making numerical fit and syntactic complexity insufficient measures of scientific credibility.
Existing approaches largely focus on improving expressions, yet the search typically retains little beyond the resulting formula and score, losing the scientific record, such as motivations and probes, that inform what to try next.
AutoSR preserves this record in a \textbf{Research State}, coupling each candidate equation with the reasoning, computational evidence, and independent review developed along its branch.
Proposer--reviewer agents develop these states under progressive-widening Monte Carlo tree search (PW-MCTS), which allocates computation across competing investigations, while the accumulated research record is ultimately synthesized into a final report that explains the leading relation and the basis for its selection.
Across nine selected challenges from two benchmark suites, AutoSR recovers algebraically equivalent relations in every case, including three cp3-bench problems that no published system recovers and six structurally diverse LSR-Transform problems.
Overall, AutoSR extends symbolic regression from equation-level search toward automated scientific investigation, allowing scientific knowledge and accumulated evidence to shape both what is explored and how the resulting equation is justified.
\end{abstract}

\section{Introduction}

% 杨老师说
% 必须要有 "自动化" 这个词, 因为很多人不想看细节, 看不懂细节, 就想知道是不是自动的. 要写上 "fully automated"
% 要说 "我们首次提出了 xxx 范式" "我们首次提出了 xxx 问题"
% "模拟人类(普朗克)的思考"
% 写作还可以参考 https://www.tandfonline.com/doi/full/10.1080/01621459.2025.2510000 发统计的四大

Scientific discovery from data seeks a relation that accounts for the phenomenon and remains useful beyond the observed sample rather than merely a formula that reproduces the observations.
Finite data underdetermine the generating function: infinitely many distinct functions can interpolate the same finite dataset \citep{bongard2007automated,cornelio2023aidescartes}.
In the presence of noise, structurally different expressions can attain nearly indistinguishable errors over the observed data while differing sharply in scientific implications such as extrapolation, dimensions, monotonicity, limiting behavior, and mechanism \citep{lacava2021srbench,matsubara2024rethinking}.
Conventional symbolic regression (SR) makes this underdetermined problem of equation discovery from finite observations tractable by searching for explicit mathematical expressions and ranking them by numerical fit and generic measures of syntactic complexity \citep{koza1992genetic,schmidt2009distilling,cranmer2023interpretable}.
These criteria favor concise candidate laws, but they do not by themselves establish scientific credibility. 
Scientists use \emph{scientific priors}---including general knowledge of the physical world, domain theory, and problem-specific requirements---to judge which hypotheses are plausible. 
Some requirements, such as dimensional consistency, monotonicity, or a boundary value, can be compiled into specialized constraints or search spaces \citep{kronberger2022shape,tenachi2023units,cornelio2023aidescartes}, whereas other forms of scientific knowledge, including causal expectations, ontology, and mechanistic plausibility, are difficult to express in a fixed grammar or scalar objective.
Moreover, scientific knowledge guides not only which equation to accept but also what to investigate next. For example, a suspicious asymptote may call for a numerical limit test, whereas an unexpected functional form may motivate comparison with a theoretical model. 
Recent LLM-based and agentic SR methods use contextual knowledge, computational tools, diagnostics, and reusable memories to improve equation proposals \citep{shojaee2024llmsr,srscientist2025,li2026restart,pang2026deliberate,zhao2026asr}.
Their advances motivate a complementary question: how can the complete motivation, experiments, failed alternatives, diagnostics, and criticism associated with each candidate become persistent objects of global search rather than transient context for producing the next equation?
% To address this gap, we introduce \textbf{Automatic Symbolic Regression (AutoSR)}, a fully automated agentic system that incorporates scientific priors by searching over research records rather than isolated expressions, and synthesizes the resulting investigation into an auditable report of the leading equation, its evidence and limitations, the explored search landscape, and credible alternatives.
To address this gap, we introduce \textbf{Automatic Symbolic Regression (AutoSR)}, a fully automated agentic system that synthesizes the resulting investigation into an auditable report that presents the leading equation with reasoning and investigation using scientific priors throughout hypothesis proposal, testing, and criticism while searching over persistent research records rather than isolated expressions.

AutoSR targets the discovery of a scientifically credible mathematical relation under a given problem specification and computation budget. 
The problem specification combines the observed data with the scientific priors used to judge which relations are plausible, including any explicit requirements that an acceptable relation should satisfy.
To preserve the broader investigation, AutoSR represents each research attempt as a \textbf{Research State}, a persistent search unit comprising a candidate equation and the broader research record of the attempt that produced it.
Research States form a branching search space in which a new state can either initiate an investigation or extend an existing one by inheriting its accumulated findings and reviews, while sibling branches pursue competing explanations independently. 
With the Research State as the search unit, AutoSR must repeatedly determine which investigation to initiate or extend next from the accumulated scientific evidence and remaining computation budget. 
We formulate this sequential decision problem as \textbf{Research-State Search}. AutoSR performs Research-State Search using Monte Carlo tree search (MCTS) \citep{coulom2006efficient,kocsis2006bandit,browne2012survey}, which allocates the computation budget across the branching search space by balancing the extension of promising investigations with the exploration of alternative branches.
Searching over complete scientific investigations rather than isolated expressions defines the broader paradigm of \textbf{Research-Space Symbolic Regression}, which AutoSR instantiates through Research-State Search.
This design provides five practical benefits:
\begin{itemize}

\item \textbf{Knowledge-guided investigation.}
Scientific priors supplied through the problem specification guide not only which candidate equations are accepted, but also how hypotheses are proposed, computationally tested, and criticized throughout the search.

\item \textbf{Accumulated research memory.}
Each Research State preserves its findings, failed attempts, and reviewer feedback, allowing descendant states along the same branch to build on accumulated evidence instead of restarting from only an equation and its score.

\item \textbf{Independent alternatives.}
By organizing Research States into separate branches, AutoSR allows alternative explanations to develop independently without being prematurely reduced to a single expression.

\item \textbf{Fully autonomous investigation.}
By combining Research-State Search with automated hypothesis generation, computational testing, and independent review, AutoSR can repeatedly select and carry out the next investigation. Once the problem specification and computation budget are provided, the system can sustain this complete process without continuous human management.

\item \textbf{Auditable scientific results.}
Because the investigation behind each candidate is preserved, the final report can connect the leading equation to its supporting evidence, limitations, research history, and credible alternatives.

\end{itemize}
In this way, AutoSR allows accumulated scientific evidence to determine what the system investigates next, rather than allowing numerical fit alone to control the search.

AutoSR is designed around the premise that scientific equation discovery can become more reliable when a system operationalizes the functional cycle of scientific inquiry used by human scientists.
Scientists do not judge a candidate equation by its numerical fit alone; they use scientific priors to propose hypotheses, design experiments or computations to test their implications, criticize the resulting evidence, and allow what they learn to determine the next investigation. 
For example, a credible black-body radiation law had to account for the measured spectrum while remaining consistent with displacement and total-radiation laws and the limiting forms then available \citep{wien1893displacement,wien1896distribution,stefan1879relation,boltzmann1884derivation,rayleigh1900remarks}.
The investigation did not end with a fitted formula: the search for a physical account led Planck to introduce energy elements proportional to frequency \citep{planck1901law}, an idea that subsequently supported Einstein's interpretation of light quanta \citep{einstein1905lightquantum}. 
Scientific inquiry can succeed where numerical fit alone cannot because each additional empirical or theoretical test can eliminate expressions that remain numerically competitive but violate scientific evidence or requirements, while the accumulated findings guide which hypothesis should be investigated next. 
AutoSR operationalizes this causal process at a computational scale: scientific priors constrain plausible equations, Research States preserve branch-specific research records, independent review challenges unsupported conclusions, and Research-State Search directs further investigation. 
In this functional sense, AutoSR mimics the exploratory cycle of human scientists without attempting to reproduce the reasoning of any individual scientist. 
Through this cycle, AutoSR is designed to distinguish scientifically credible equations from expressions that are merely accurate over the observed data.

We evaluate AutoSR through exact equation recovery on selected challenge problems from two benchmark suites. 
On all nine selected problems, AutoSR recovers relations algebraically equivalent to the ground truth. 
These include three cp3-bench equations that were not recovered by any of twelve published SR systems \citep{thing2025cp3} and six structurally diverse LSR-Transform equations \citep{shojaee2025llmsrbench}, all recovered using the same core search procedure and agent roles. 
The present evidence establishes feasibility on these selected challenge cases; repeated-run reliability, component ablations, scheduling efficiency, and domain validation remain subjects for future evaluation.

Our contributions are threefold. 
First, we introduce Research-Space Symbolic Regression and, to our knowledge, the first formulation of Research-State Search as the sequential decision problem underlying this paradigm. 
Second, we develop AutoSR as a fully automated implementation that combines persistent Research States, independent scientific review, autonomous computational investigation, and MCTS-based search. 
Third, we demonstrate exact recovery on nine selected challenge equations spanning two benchmark suites. 
The remainder of the paper provides the related work review, defines the AutoSR method, presents the selected benchmark evaluation, and discusses the implications, limitations, and future evaluation of treating symbolic regression as automated scientific inquiry.

\section{Related Work}

\subsection{Conventional and Prior-Informed Symbolic Regression}

Symbolic regression jointly searches for the structure and parameters of an explicit mathematical expression.
Genetic programming established expression trees, evolutionary variation, and fitness-based selection as a general formulation of this search \citep{koza1992genetic}.
Early systems for scientific discovery extended expression search with active probing of dynamical systems \citep{bongard2007automated} and searches for invariant relations in experimental data \citep{schmidt2009distilling}.
Subsequent approaches span deterministic model building such as FFX \citep{mcconaghy2011ffx}, sparse identification over a prescribed function library in SINDy \citep{brunton2016discovering}, compressed-sensing feature construction in SISSO \citep{ouyang2018sisso}, probabilistic search over expression trees in Bayesian SR \citep{jin2020bayesian}, and physics-inspired decomposition by symmetry, separability, and dimensional analysis in AI Feynman \citep{udrescu2020aifeynman}.
Modern systems such as PySR and FEX combine expressive search spaces with specialized optimization \citep{cranmer2023interpretable,jiang2023fex}, while parallel symbolic enumeration shares subtree computations to evaluate very large collections of expressions efficiently \citep{ruan2026parallel}.

Across these algorithmic families, candidates are ordinarily selected by predictive error together with parsimony, sparsity, a structural prior, or a related model-selection criterion.
SRBench compares contemporary methods on both known-ground-truth and black-box tasks and shows that predictive accuracy, symbolic recovery, and expression complexity expose different aspects of performance \citep{lacava2021srbench}.
SRSD further emphasizes realistic sampling ranges, irrelevant variables, and normalized tree-edit distance for scientific-discovery evaluation \citep{matsubara2024rethinking}, while LLM-SRBench adds transformed and synthetic scientific equations together with safeguards against equation memorization \citep{shojaee2025llmsrbench}.
These benchmarks make clear that syntactic size is a useful operational proxy, but neither low error nor a short expression alone establishes that a recovered relation has the intended scientific meaning.

Prior knowledge can narrow this ambiguity when it can be formalized.
It has been incorporated through data generated from known constraints \citep{kubalik2020prior}, bounds and derivative-shape restrictions such as monotonicity or convexity \citep{kronberger2022shape}, dimensional consistency enforced during expression construction \citep{tenachi2023units}, user-specified structural hypotheses for neural SR \citep{bendinelli2023controllable}, and probabilistic priors over functions, structures, and parameters \citep{bartlett2023priors}.
AI-Descartes evaluates candidate formulas against logical background axioms \citep{cornelio2023aidescartes}, and AI-Hilbert jointly searches for polynomial laws consistent with data and background theory while producing formal certificates \citep{corywright2024aihilbert}.
AutoSR treats these formal constraints as valuable special cases: its problem specification may include them, but also admits scientific knowledge that remains in natural language and must influence hypothesis formation, computational testing, and criticism rather than a single fixed objective.

\subsection{Learned and Tree-Search-Guided Symbolic Regression}

Learned SR methods replace or augment hand-designed proposal distributions with neural priors.
Deep Symbolic Regression trains a recurrent policy with a risk-seeking objective \citep{petersen2021deep}, whereas NeSymReS and end-to-end Transformer systems pretrain on synthetic equations and infer expressions from sets of numerical observations \citep{biggio2021nesymres,kamienny2022endtoend}.
A unified deep-SR framework combines recursive simplification, neural-guided search, pretraining, genetic programming, and linear models, illustrating that learned generation and explicit search are complementary rather than mutually exclusive \citep{landajuela2022unified}.

MCTS provides a general mechanism for allocating simulations in a tree by repeatedly selecting, expanding, evaluating, and backing up nodes \citep{coulom2006efficient,browne2012survey}.
UCT applies an upper-confidence bandit rule at each internal node to balance exploration and exploitation \citep{kocsis2006bandit}, and progressive widening extends the approach to spaces in which enumerating every action at a node is impractical \citep{couetoux2011continuous}.
In SR, Symbolic Physics Learner uses MCTS to construct mathematical expression trees \citep{sun2023spl}; DGSR-MCTS uses a pretrained and online-refined neural mutation model \citep{kamienny2023dgsr}; and TPSR adds MCTS lookahead to Transformer decoding with accuracy--complexity feedback \citep{shojaee2023tpsr}.
RSRM combines expression-tree MCTS with double Q-learning and learned subtree operators \citep{xu2024rsrm}, while SR4MDL searches subformulas using a learned minimum-description-length objective \citep{yu2025sr4mdl}.
More recent methods introduce extreme-bandit allocation and nonlocal mutation or crossover actions \citep{huang2025mcts}, or search permutation-invariant expression graphs with neural and constraint guidance \citep{xiang2025gsr}.
Thus MCTS itself, and its use in SR, are established; the distinction in AutoSR is the state being searched and evaluated.
Previous SR applications assign nodes to tokens, subexpressions, formulas, or expression graphs, whereas one AutoSR expansion produces a persistent proposer--experiment--review record whose value and ancestry determine the allocation of subsequent investigations.

\subsection{LLM-Guided and Agentic Symbolic Regression}

LLMs introduce scientific language and program synthesis as additional proposal mechanisms.
In-Context Symbolic Regression iteratively refines LLM-generated formulas with external coefficient optimization \citep{merler2024icsr}, and LLM4ED uses LLMs as black-box optimizers and evolutionary operators for equation discovery \citep{du2024llm4ed}.
The Scientific Generative Agent couples LLM proposals with simulation and continuous optimization in a bilevel loop \citep{ma2024bilevel}.
LaSR evolves a natural-language concept library alongside symbolic hypotheses \citep{grayeli2024lasr}, while LLM-SR generates programmatic equation skeletons inside a multi-island evolutionary search using scientific context \citep{shojaee2024llmsr}.
RAG-SR instead trains an online language model and retrieves previously searched expressions to guide feature construction without large-scale pretraining \citep{zhang2025ragsr}.

Recent work moves from LLM proposal components toward longer-horizon agents.
SR-Scientist gives an agent tools for data analysis and equation evaluation and preserves strong equations in an experience buffer \citep{srscientist2025}.
RESTART uses residual diagnostics for short-term structural correction and distills successful refinements into a long-term structure library \citep{li2026restart}; Deliberate Evolution separates equation generation from search guidance through adaptive operators, diagnostic tools, and reflective trajectory memory \citep{pang2026deliberate}; and the Iterated Agent uses natural-language rationales as semantic operators in an evolutionary search \citep{song2025iterated}.
Other concurrent preprints let an LLM control a conventional SR engine \citep{xie2026controlling}, combine granular influence feedback with term generation and MCTS \citep{saveliev2026influence}, or coordinate Generator, Analyst, Simplifier, and Reviewer roles through process memory \citep{zhao2026asr}.
These systems already go substantially beyond a formula--score loop, so AutoSR does not claim the first use of tools, diagnostics, memory, multiple roles, or iterative scientific reasoning in SR.
Its narrower contribution is to make each complete, artifact-rich investigation a persistent node of global search, keep its memory local to an ancestral branch, append a separate review to the node before reward backup, and synthesize the resulting tree as auditable evidence rather than using the trace only to improve subsequent equation proposals.

Tree search over richer work products also appears in general research agents: AIDE searches code-solution states \citep{jiang2025aide}, and AI Scientist-v2 searches branching experimental workflows \citep{yamada2025aiscientist2}.
AutoSR specializes this broader direction to scientific equation discovery by defining the contents, inheritance, independent review, and search semantics of a Research State.

\section{Method}

\subsection{Problem Formulation and System Overview}

AutoSR receives a problem specification $\mathcal{P}$ and a computation budget $B$.
The specification defines the scientific question, the observed data, the meanings of the variables, the available supporting evidence, the evaluation criterion, and any requirements that an acceptable relation should satisfy.
Given $(\mathcal{P},B)$, the system returns not only a leading relation $\hat f$, but a final report $\mathcal{R}_{\mathrm{final}}$ that connects this relation to its evidence, limitations, research history, and credible alternatives:
\begin{equation}
\operatorname{AutoSR}(\mathcal{P},B)
\longrightarrow
(\hat f,\mathcal{R}_{\mathrm{final}}).
\label{eq:autosr-objective}
\end{equation}
The task is therefore broader than minimizing a fixed loss over an expression space.
AutoSR must use the available scientific information to produce, test, criticize, and compare candidate relations while deciding how to allocate the remaining budget.

The concrete interface to $\mathcal{P}$ consists of \texttt{problem.md}, data, and supporting files.
The supporting material may include background knowledge, figures, documents, existing code, and explicit scientific requirements.
Requirements such as dimensional consistency, monotonicity, convexity, asymptotic behavior, special values, and known theoretical limits can often be checked directly \citep{kronberger2022shape,tenachi2023units,cornelio2023aidescartes}.
Other priors, including causal expectations, ontology, and mechanistic plausibility, may remain expressed in scientific language rather than in a fixed grammar or penalty.
AutoSR supplies the complete specification to both its proposal and review roles, allowing formal requirements and broader scientific knowledge to influence hypothesis generation, computational testing, and criticism.

\begin{figure}[t]
\centering
\includegraphics[width=\linewidth]{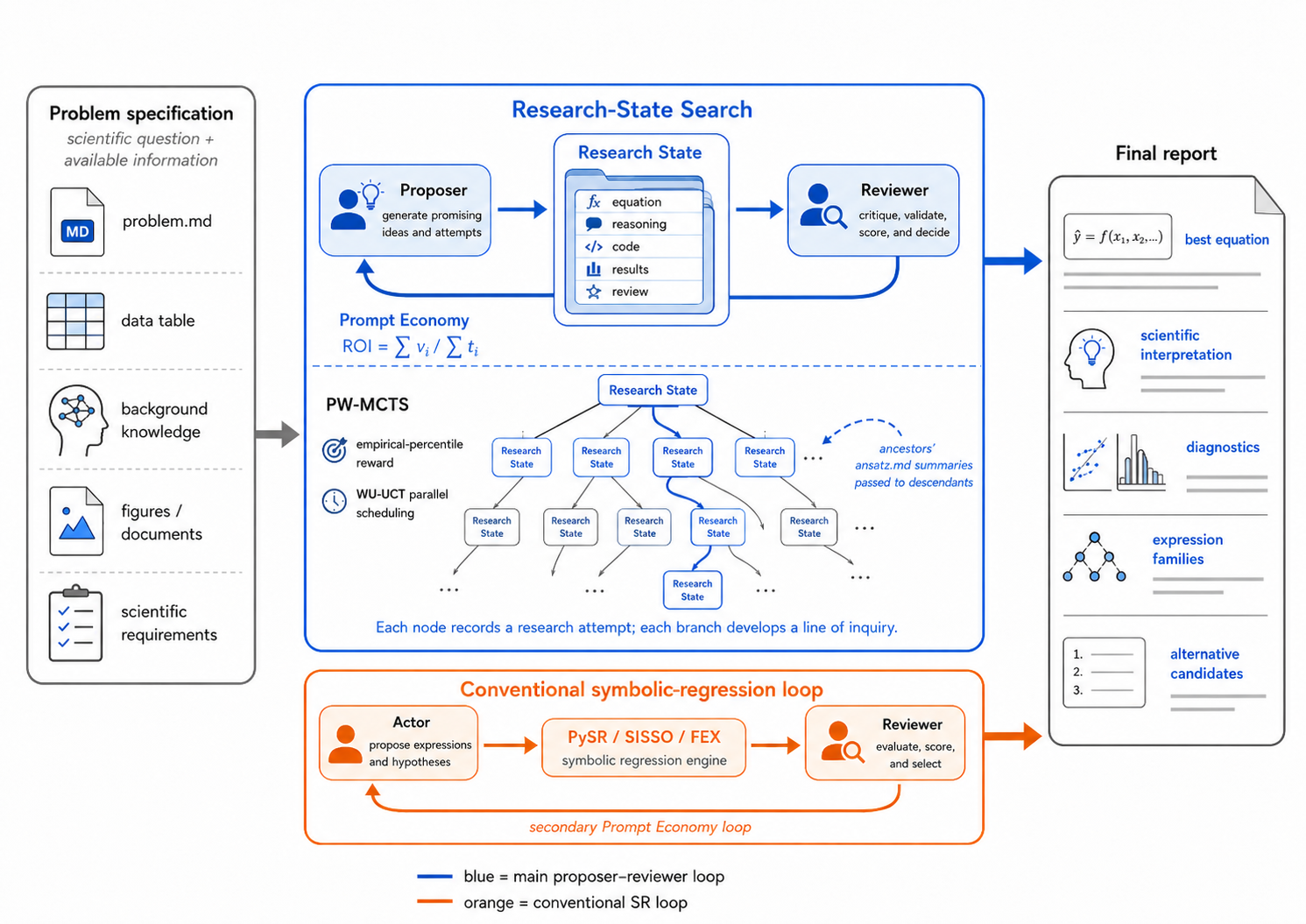}
\caption{AutoSR uses a proposer--reviewer loop to build a tree of Research States with progressive-widening MCTS.
The problem specification supplies data, scientific context, and explicit requirements to the investigation.
A secondary actor--reviewer loop advances conventional symbolic regression.
The Prompt Economy annotation uses the AutoSR-specific reuse metric defined in Equation~\ref{eq:prompt-economy-roi}.
The final report combines the strongest candidates, their supporting evidence, and the resulting search landscape.}
\label{fig:autosr-overview}
\end{figure}

Starting from a root that contains $\mathcal{P}$, AutoSR builds a tree of Research States.
Each tree expansion executes an automated proposer--reviewer cycle, and the search controller selects which branch to initiate or extend next.
Progressive-widening Monte Carlo tree search (PW-MCTS) \citep{browne2012survey,couetoux2011continuous} balances new lines of inquiry against continuations of promising branches, while a pending-aware UCT rule \citep{kocsis2006bandit,liu2020watch} permits multiple long-running expansions to proceed asynchronously.
A complementary loop applies established symbolic-regression packages, and the final synthesis compares the strongest candidates from both processes.
Figure~\ref{fig:autosr-overview} summarizes this workflow.

\subsection{Research States and Branch-Specific Memory}

A Research State is the persistent record produced by one research attempt.
For a non-root node $i$, write
\begin{equation}
\mathcal{S}_i
=
(f_i,m_i,\mathcal{A}_i,\rho_i,s_i;p_i),
\label{eq:research-state}
\end{equation}
where $f_i$ is the candidate relation, $m_i$ is its scientific motivation and reasoning, $\mathcal{A}_i$ contains the computational artifacts and findings, $\rho_i$ is the review record, $s_i$ is the reviewer score, and $p_i$ identifies the parent Research State.
The artifacts may include code, execution results, fitted parameters, figures, residual analyses, diagnostics, and unsuccessful tests.
The scalar $s_i$ supports search control, whereas the full record $(m_i,\mathcal{A}_i,\rho_i)$ preserves the evidence needed to interpret that score.

The central artifact in each state is \texttt{ansatz.md}.
It serves as both summary and index: it states the current hypothesis, condenses the main argument and results, records the review, and points to the supporting files.
The local contents of $\mathcal{S}_i$ record attempt $i$, while the ordered path from the root to $i$ forms the memory of the corresponding line of inquiry.
When a proposer extends state $i$, it receives the \texttt{ansatz.md} files on this ancestral path in order.
It does not receive summaries from sibling branches.
A child of the root therefore begins an independent investigation, whereas a deeper child can build on the useful ideas, failed trials, and criticism accumulated by its ancestors.
This separation allows competing explanations to develop independently without discarding the research history within any one branch.

\subsection{Automated Investigation and Independent Review}

The proposer receives $\mathcal{P}$ together with the branch memory selected by the search controller.
It develops an ansatz, writes and executes code, fits parameters, studies plots and numerical behavior, and revises the candidate in response to the resulting evidence.
The proposer may change transformations, objectives, optimizers, or computational methods when the current formulation is unstable or scientifically misleading, and it may call symbolic-regression packages or other scientific tools when useful.
These decisions are not prescribed as a fixed pipeline: they are part of the investigation recorded in $(m_i,\mathcal{A}_i)$.

After the proposal is complete, a separate reviewer role reads the problem specification, \texttt{ansatz.md}, and the supporting artifacts.
It reruns key computations, checks the stated evaluation criterion, examines residuals, and searches for leakage, hidden parameters, violated requirements, or unsupported conclusions.
Held-out data and hidden evaluation instructions can be supplied through a reviewer-only channel so that the proposer cannot optimize directly against them.
The reviewer records its diagnosis $\rho_i$ and score $s_i$ in \texttt{ansatz.md}.
A mandatory scientific requirement remains an acceptance condition rather than a tradeable improvement in numerical fit.
The scalar score is used to allocate search effort; it does not replace the reviewer’s evidence or the scientific comparison preserved in the final report.
Appending this review completes $\mathcal{S}_i$, after which its descendants can inherit both the proposal and its criticism.

Producing a Research State can require repeated reasoning, coding, execution, and revision.
AutoSR draws on Prompt Economy, which treats the maintenance of role prompts and handoff protocols as a bounded engineering cost and favors reusing a stable prompt surface across many useful invocations \citep{perspectivegap2026,agon2026}.
We operationalize this principle for AutoSR with a time-based, system-specific reuse metric.
Let $t_i$ be the time spent writing or revising prompt $i$, and let $v_i$ be the number of times that prompt is executed:
\begin{equation}
\mathrm{ROI}_{\mathrm{AutoSR}}
=
\frac{\sum_i v_i}{\sum_i t_i}.
\label{eq:prompt-economy-roi}
\end{equation}
This AutoSR-specific quantity adapts the Prompt Economy principle rather than reproducing the ROI formula in the cited work; it is a system-design criterion, not the scientific objective in Equation~\ref{eq:autosr-objective}.
Reusing the same proposer and reviewer roles at every expansion increases the number of autonomous investigations without requiring a new prompt for each candidate or domain.

\subsection{Research-State Search}

Research-State Search is the sequential decision problem of choosing which investigation to initiate or extend under the remaining budget.
At scheduling step $t$, let $\mathcal{T}_t$ denote the tree of created Research States, including completed states and expansions still in flight.
Each completed proposer--reviewer cycle supplies one new node and one observed reward; in this sense, a complete research attempt plays the role of an MCTS evaluation \citep{coulom2006efficient,browne2012survey}.
AutoSR allocates these evaluations with progressive-widening MCTS (PW-MCTS; \citealp{couetoux2011continuous}).

For node $i$, let $C_i$ be its set of created children, $N_i$ the number of completed evaluations whose selected paths pass through $i$, and $O_i$ the corresponding number of evaluations that have been dispatched but have not yet returned.
The controller may create another child of $i$ when

\begin{equation}
|C_i|
<
k_{\mathrm{pw}}\max(N_i+O_i,1)^{\beta}.
\label{eq:progressive-widening}
\end{equation}
Here $k_{\mathrm{pw}}>0$ controls the overall width and $0<\beta<1$ makes the allowed branching factor grow sublinearly with the number of completed and pending evaluations.
If the condition holds, the proposer creates a new child from the branch memory of $i$; otherwise, the controller continues through an existing child according to the tree policy below.
Progressive widening therefore admits more independent or descendant investigations as evidence accumulates without allowing early branching to consume the entire budget.

The reviewer score can depend on the evaluation criterion, scientific requirements, and the supporting analysis, and its raw scale may differ across problems.
Let $\mathcal{V}_t$ be the set of Research States completed by scheduling step $t$, with larger $s_i$ indicating a stronger review.
AutoSR converts each score to its empirical percentile within the current completed set:
\begin{equation}
r_i(t)
=
10\times
\frac{1}{|\mathcal{V}_t|}
\sum_{\ell\in\mathcal{V}_t}
\mathbf{1}[s_{\ell}\leq s_i].
\label{eq:percentile-reward}
\end{equation}
Percentile rewards lie in $(0,10]$ and are interpreted at the current scheduling step because their reference set grows as reviews return.
This common rank scale prevents the magnitude or units of a task-specific evaluator from changing the exploration pressure.
For a child $j$, let $Q_j(t)$ be the mean percentile reward of the completed Research States in the subtree rooted at $j$.

Because a proposer--reviewer cycle may run for minutes or hours, scheduling evaluations only after earlier calls finish would leave substantial parallel capacity unused.
AutoSR therefore adapts Watch the Unobserved in UCT (WU-UCT; \citealp{liu2020watch}) to Research-State Search.
Among existing children, the controller selects according to
\begin{equation}
U_t(i,j)
=
Q_j(t)
+
c\sqrt{
\frac{\log\max(N_i+O_i,1)}{N_j+O_j}
}.
\label{eq:wu-uct}
\end{equation}
The in-flight counts allow dispatched work to affect both selection and progressive widening before its reviewer score becomes observable.
They discourage redundant simultaneous expansion of the same branch while retaining the exploitation term supplied by completed reviews.

One asynchronous search iteration consists of four operations.
First, the controller traverses the tree from the root, applying Equation~\ref{eq:progressive-widening} to decide whether to create a child and Equation~\ref{eq:wu-uct} when it must continue through existing children.
Second, before dispatching the selected proposer--reviewer cycle, it increments $O_i$ along the selected path so that later scheduling decisions observe the pending work.
Third, the proposer and reviewer produce the new Research State without blocking other dispatched evaluations.
Finally, when the review returns, the controller decrements the path's in-flight counts, increments its completed counts, adds the new score to $\mathcal{V}_t$, and updates the percentile rewards and subtree means used by subsequent selections.
This cycle continues until budget $B$ is exhausted.
The controller then stops dispatching new work, resolves the remaining in-flight evaluations, and collects the completed Research States for final synthesis.

\subsection{Conventional Symbolic-Regression Loop}

A secondary Prompt Economy loop applies conventional symbolic regression to the same problem specification.
This loop complements the flexible investigations in the Research-State tree with mature expression-search engines and their specialized optimizers.
Its actor configures and runs methods such as PySR \citep{cranmer2023interpretable}, SISSO \citep{ouyang2018sisso}, and FEX \citep{jiang2023fex}, then studies their equations, residuals, and failure modes.
Its reviewer examines the search space and returned candidates, identifies overlooked transformations or settings, and directs the next package run.
One package execution is therefore an experiment within a sustained actor--reviewer process rather than the complete conventional-SR contribution.
The loop passes its strongest equations, diagnostics, and reviews to the same final synthesis used for Research-State candidates.

\subsection{Final Selection and Report Synthesis}

After both search processes end, AutoSR ranks completed candidates by reviewer score while retaining the corresponding requirement checks and review records.
A candidate that violates a requirement identified by the problem specification cannot be presented as an acceptable scientific relation solely because it attains a strong numerical fit.
The leading eligible candidate becomes $\hat f$ in Equation~\ref{eq:autosr-objective}, but ranking determines only which result the report presents first; it does not discard competitive alternatives or their evidence.

The first part of $\mathcal{R}_{\mathrm{final}}$ explains the leading result through its formula, measured performance, scientific interpretation, diagnostics, and limitations.
The second reconstructs the Research-State search landscape: it presents a report-ready tree, groups states into expression families, summarizes the evidence associated with each family, and traces clear evolution paths through the parent relation in Equation~\ref{eq:research-state}.
An appendix presents the next four candidates as standalone alternatives.
When several relations remain competitive, the report compares their empirical and scientific implications instead of forcing an unsupported single conclusion.
The equation remains the central result, while the preserved artifacts and reviewer records make the investigation behind that result available for audit.

\section{Experiments}

The present evaluation asks whether AutoSR can recover exact relations on selected equations that pose substantial structural challenges to existing symbolic-regression systems.
We report nine recovery cases drawn from two benchmark suites.
Because the problems were selected for difficulty and one recovered trajectory is reported for each task, these experiments demonstrate capability on the selected cases rather than benchmark-wide accuracy or stochastic reliability.

\subsection{Evaluation Scope and Protocol}

The first challenge set comes from cp3-bench \citep{thing2025cp3}, which evaluates twelve symbolic-regression methods on twenty-eight problems in cosmology and astroparticle physics.
We selected C3g, C5f, and C6c because none of the twelve methods recovered their ground-truth relations.
The three tasks cover dependent input features, a single expression that unifies cusp and core density profiles, and a strongly oscillatory gravitational-wave signal with an additional mass parameter.

The second challenge set comes from LSR-Transform in LLM-SRBench \citep{shojaee2025llmsrbench}.
LSR-Transform contains 111 equations obtained by solving established scientific relations for unusual target variables.
The best official symbolic accuracy is 31.53\%.
We selected six tasks that cover complementary structures: nested radicals, trigonometric inversion, fractional powers, paired exponentials, a high-dimensional rational expression, and logarithmic inversion.
The selection was based on structural complexity and was completed before we examined public per-task results from other systems.

Every task uses the same core search procedure and role prompts.
The proposer receives the task description and training data, while the ground-truth equation remains hidden until the search ends.
Internet access is prohibited.
An external evaluator checks algebraic equivalence after the run.
Reported times measure only the proposer sessions that produced the recovered relation.
They exclude other explored branches, reviewer calls, parallel work, and conventional-SR runs, and therefore should not be interpreted as the total computational cost of AutoSR.

\subsection{Selected cp3-bench Recovery}

AutoSR recovered a relation algebraically equivalent to the ground truth on each of the three selected cp3-bench problems, where every published baseline failed to recover the corresponding analytical relation.

\begin{table}[ht]
\centering
\small
\caption{Results on the selected cp3-bench problems.
MSE values are reported to two significant digits.}
\label{tab:cp3-results}
\begin{tabularx}{\linewidth}{@{}l Y Y r@{}}
\toprule
Problem & Best cp3-bench result & AutoSR result & Proposer time \\
\midrule
C3g & QLattice; MSE $4.7\times10^{-6}$; ground truth not recovered & Algebraically equivalent; MSE $9.0\times10^{-19}$ & 8 min 58 s \\
C5f & GPG; MSE $3.3\times10^{-5}$; ground truth not recovered & Algebraically equivalent; MSE $6.2\times10^{-32}$ & 15 min 44 s \\
C6c & GPG; MSE $1.1\times10^{-1}$; ground truth not recovered & Algebraically equivalent; MSE $6.1\times10^{-30}$ & 33 min 1 s \\
\bottomrule
\end{tabularx}
\end{table}

All three recovered formulas simplify to the known analytical relations.
Their surface forms differ from the benchmark expressions, which is consistent with nontrivial algebraic reconstruction but does not by itself rule out the influence of memorized scientific knowledge.
On three problems where twelve established systems recovered zero relations, AutoSR recovered three.

\subsection{Selected LSR-Transform Recovery}

The LLM-SRBench paper reports aggregate results but does not release the original baselines' predictions for individual LSR-Transform tasks.
We therefore report AutoSR's per-task results without claiming a direct per-task comparison with the published baselines.

\begin{table}[ht]
\centering
\footnotesize
\caption{Results on the selected LSR-Transform problems.
NRMSE values are reported to two significant digits.}
\label{tab:lsr-transform-results}
\begin{tabularx}{\linewidth}{@{}l Y r@{}}
\toprule
Problem & AutoSR result & Proposer time \\
\midrule
\texttt{I.32.17\_4\_3} & Algebraically equivalent; NRMSE $1.8\times10^{-7}$ & 16 min 13 s \\
\texttt{I.29.16\_0\_0} & Algebraically equivalent; NRMSE $1.2\times10^{-7}$ & 12 min 32 s \\
\texttt{II.6.15a\_2\_0} & Algebraically equivalent; NRMSE $1.0\times10^{-7}$ & 7 min 47 s \\
\texttt{II.35.18\_0\_0} & Algebraically equivalent; NRMSE $3.6\times10^{-7}$ & 6 min 26 s \\
\texttt{II.36.38\_3\_0} & Algebraically equivalent; NRMSE $8.0\times10^{-8}$ & 9 min 7 s \\
\texttt{III.4.33\_3\_0} & Algebraically equivalent; NRMSE $7.2\times10^{-8}$ & 8 min 9 s \\
\bottomrule
\end{tabularx}
\end{table}

AutoSR recovered an algebraically equivalent relation on all six tasks.
The six equations require substantially different transformations, while the search procedure and agent roles remain unchanged.

\subsection{Scope of the Present Evidence}

Taken together, the nine cases show that AutoSR can recover algebraically exact relations across several difficult expression structures while preserving a common search procedure and set of agent roles.
They do not estimate a benchmark-wide recovery rate, the probability of success under repeated runs, or the contribution of any individual system component.
The proposer times in Tables~\ref{tab:cp3-results} and~\ref{tab:lsr-transform-results} also do not measure total system cost.
Accordingly, the present results should be interpreted as selected challenge-case demonstrations rather than a comprehensive comparison with existing symbolic-regression systems.
Section~\ref{sec:future-evaluation} specifies the controlled and domain evaluations required to extend this evidence.

\section{Discussion}

The significance of the selected recoveries lies in the common research workflow that produced them.
Across the selected problems, AutoSR retained the same core search procedure and agent roles.
These cases show that a common sustained research workflow can recover structurally diverse equations.

The central distinction is the unit over which the system searches.
A conventional expression-level search can retain only a formula and its score, whereas a Research State is required to preserve the motivation, computations, failed tests, and criticism associated with one attempt.
This persistent record allows evidence to influence what is investigated next, while branch-specific memory permits competing explanations to develop independently.
Within this structure, scientific priors can guide not only which relation is accepted, but also which hypothesis is proposed, which implication is tested, and which branch receives further computation.

Persistent Research States also support a final report that connects the leading relation to its evidence, limitations, research history, and credible alternatives.
In this setting, \emph{fully automated} means that AutoSR can continue the investigation after receiving a problem specification and computation budget; it does not remove the scientist's role in formulating the problem or interpreting the result.
AutoSR therefore provides a concrete implementation of Research-Space Symbolic Regression in which the search process retains the broader investigation rather than only its final equation.

\section{Limitations and Future Evaluation}
\label{sec:future-evaluation}

The present evidence is limited to nine selected challenge cases, with one recovered trajectory reported for each.
It does not establish benchmark-wide accuracy, repeated-run reliability, or the computation required to obtain a recovery with a specified probability.
The reported proposer times also exclude reviews, unsuccessful and parallel branches, conventional-SR runs, token usage, and orchestration overhead.
Future evaluation should therefore report repeated independent runs, success at fixed budgets, total model calls and tokens, completed Research States, wall-clock time, concurrency, and time to first recovery.

The experiments do not isolate the source of the successful recoveries.
Matched-budget ablations should compare direct LLM generation, a single tool-using agent, a proposer--reviewer loop without tree search, flat best-of-$N$ Research States, and the complete system.
The conventional-SR loop should be removed in a complementary ablation, and a supplied-context ablation should compare data and variable names alone with the full scientific background and requirements.
These comparisons would separate the effects of tool use, review, branch-specific memory, Research-State Search, and explicitly supplied scientific context.

Exact benchmark recovery also does not establish that AutoSR can distinguish scientifically credible relations from numerically competitive alternatives under noise, extrapolation, incomplete theory, or conflicting evidence.
Nor do the current experiments assess the final reports, alternative relations, or preserved search landscapes.
Evaluation on newly constructed problems and private domain data is needed because memorized scientific knowledge cannot be ruled out on established benchmarks, particularly when pretrained models are used \citep{matsubara2024rethinking,shojaee2025llmsrbench}.
Separate proposer and reviewer roles also provide procedural separation without guaranteeing independent judgment when both rely on similar pretrained knowledge.
External computational checks and domain-expert review will therefore remain necessary.

The search scheduler requires direct evaluation as well.
Comparisons of raw reviewer scores with empirical-percentile rewards and sequential UCT with WU-UCT should measure result quality at fixed computation, redundant concurrent branch selection, worker utilization, and time to the best candidate.
Ongoing studies in materials science, mathematical biology, and geotechnical engineering will provide the test of whether AutoSR can use problem-specific requirements to produce scientifically useful relations.

\paragraph{Code availability.}
The implementation of AutoSR is available in the \href{https://github.com/AutoResearch-Factory/AutoSR}{AutoResearch-Factory/AutoSR}.

\begingroup
\small
\bibliographystyle{alpha}
\bibliography{sample}
\endgroup

\end{document}